\documentclass{svproc}
\usepackage{url}
\usepackage{graphicx}

\begin{document}
\mainmatter              
\title{A General-Purpose Device for Interaction with LLMs}
\titlerunning{A General-Purpose Device for Interaction with LLMs}  
%
\author{Jiajun Xu \inst{1} \and Qun Wang\inst{2} \and
Yuhang Cao\inst{3} \and Baitao Zeng\inst{3} \and Sicheng Liu\inst{2}}
\authorrunning{Jiajun Xu et al.} 
%
\tocauthor{Jiajun Xu, Qun Wang,
Yuhang Cao, Baitao Zeng, Sicheng Liu}
\institute{Department of Electrical and Computer Engineering, University of Southern California, Los Angeles, CA, 90089,\\
\email{jiajunx@usc.edu}
\and
Department of Computer Science, San Francisco State University, San Francisco, CA, 94132,\\
\email{qunwang@sfsu.edu, sliu24@mail.sfsu.edu}
\and
InnoGen AI, Palo Alto, CA, 94306,\\
\email{suqianxi@mail.ustc.edu.cn, helloworld.zbt.2022@gmail.com}
}

\maketitle              

\begin{abstract}
This paper investigates integrating large language models (LLMs) with advanced hardware, focusing on developing a general-purpose device designed for enhanced interaction with LLMs. Initially, we analyze the current landscape, where virtual assistants and LLMs are reshaping human-technology interactions, highlighting pivotal advancements and setting the stage for a new era of intelligent hardware. Despite substantial progress in LLM technology, a significant gap exists in hardware development, particularly concerning scalability, efficiency, affordability, and multimodal capabilities. This disparity presents both challenges and opportunities, underscoring the need for hardware that is not only powerful but also versatile and capable of managing the sophisticated demands of modern computation. Our proposed device addresses these needs by emphasizing scalability, multimodal data processing, enhanced user interaction, and privacy considerations, offering a comprehensive platform for LLM integration in various applications. 
\end{abstract}
\section{Introduction}

In recent years, Virtual Assistants (VAs) such as Amazon's Alexa, Apple's Siri, Google Assistant, and Microsoft's Cortana have become integral to our daily lives, facilitating a range of services easily. Despite their widespread adoption, traditional VAs often struggle with processing complex commands and providing accurate responses. The recent emergence of Large Language Models (LLMs) like ChatGPT and Claude provide solutions to overcome these limitations and promise a new era of Intelligent Assistants (IAs) that are capable of interpreting intricate contexts and delivering more satisfactory responses.


The popular trend of IAs reflects a growing demand for automation in both professional and personal spheres. These advanced assistants are designed to navigate the complexities of various scenarios, from corporate environments to household tasks, offering a seamless interaction experience.
Most IAs are implemented on smartphones, including AutoDroid \cite{autodroid}, GptVoiceTasker \cite{gptvoice}, and EdgeMoE \cite{edgemoe}.
AutoDroid integrates LLMs for task automation on Android devices, enhancing task execution without manual input by combining app-specific insights with LLMs' general knowledge. It has shown superior task automation capabilities, significantly outperforming previous methods in accuracy and efficiency \cite{autodroid}.
GptVoiceTasker used LLMs to boost mobile task efficiency and user interaction. It learns from past commands to improve responsiveness, speeding up task completion \cite{gptvoice}.

The integration of LLMs into smartphone applications demonstrates the ability of LLMs to automate complex tasks efficiently, even on devices with limited computing and memory resources. These developments highlight a significant shift toward more intelligent, responsive, and user-centric mobile applications driven by LLM technology. However, the reliance on smartphones restricts their ability to process multidimensional inputs and hinders seamless integration with existing infrastructures. To truly harness the potential of IAs, there is a pressing need for a novel framework that synergizes both software and hardware components. This approach would enhance the capability of IAs to understand and execute complex commands and enable their application across a wider range of contexts, thereby revolutionizing the way we interact with technology in our daily lives.

To fully harness the capabilities of IAs as agents that control and interact with diverse applications, it is crucial to deploy them on independent hardware platforms tailored to LLM requirements and specific application needs. This approach would facilitate the modularization of system components and the creation of well-defined APIs that are easy to implement, thereby accelerating development. There is limited research on this topic.

LLMind employs LLMs to manage domain-specific AI modules and IoT devices. It facilitates complex task execution by enabling natural user interactions through a social media-like platform. Here, LLMs generate execution plans via control scripts derived from language-code transformations using finite-state machines. By incorporating semantic analysis and response optimization, LLMind can improve IoT control, enhance user experience, and foster an evolving, intelligent IoT ecosystem \cite{llmind}.
EdgeMoE offers a distributed framework tailored for micro-enterprises, enabling the partitioning of LLMs across devices according to their computational capabilities. This strategy supports parallel processing, enhances privacy, and reduces response times by leveraging the specific hardware capabilities of small businesses \cite{pipline}.

However, there is still a huge demand to accurately process multidimensional inputs while guaranteeing the response delay and preserving users' privacy. 
Therefore, we designed a novel edge caching-enabled IA framework to solve the above challenges.
Specifically, our contribution to this work is summarized as follows:
We introduce a novel framework designed to bridge the gap between input from edge devices and LLMs data process center. Moreover, our framework is uniquely tailored to integrate various forms of input, with a special focus on enhancing voice inputs through local preprocessing to achieve more accurate input information. We integrated local caching to facilitate rapid processing and enable computational reuse, significantly improving system efficiency and reducing response times. This comprehensive approach addresses critical challenges in edge computing and sets a new standard for efficient model interaction with edge devices.
The remainder of this paper is organized as follows. Section 2 will discuss the proposed framework design. Section 3 discusses edge device implementation. Section 4 discusses the future work and challenges. Section 5 provides a paper's conclusion.


\section{Framework of Proposed Design}
In this section, we will first discuss the user demands for LLM-enabled IA and the goals that our proposed framework wants to achieve.
\subsection{User Needs and Design Goals}

Considering the advancements in LLM integration and the specifications outlined in our hardware design, we aim to achieve the following design goals.

   Our primary goal is to develop a general-purpose device that harmoniously integrates both hardware and software components with LLMs. This device will enable intuitive voice interactions, leveraging the sophisticated language processing capabilities of LLMs.
    By focusing on a seamless connection between the hardware and the LLM, we aim to create a device that can interpret and respond to user commands accurately and efficiently, enhancing the overall user experience.


    Moreover, accessibility is important to enable IA to support critical applications. We aim to create a device that is not only affordable but also easily integrated with existing platforms and technologies. This involves designing the device so that it can be adopted in various environments—from personal homes to businesses—without requiring significant changes to existing infrastructures or incurring high costs.


     Our device is designed to handle multi dimensional inputs, including audio, video, and data from various environmental sensors. The integration of these diverse input methods enables the device to handle complex tasks that require a deeper understanding of the context and environment, thereby providing more accurate and relevant responses and actions.

Our design goals are aligned with the evolving landscape of IA. By focusing on creating a general-purpose, affordable, and multi modal-capable device, we aim to address the current limitations in user interaction paradigms and set a new standard for intelligent device interactions. This approach not only aims to enhance user productivity and experience but also ensures that our device remains adaptable and relevant in the rapidly progressing field of LLM and AI technology.

\subsection{Framework Overview}

\begin{figure}
    \centering
    \includegraphics[width=\linewidth]{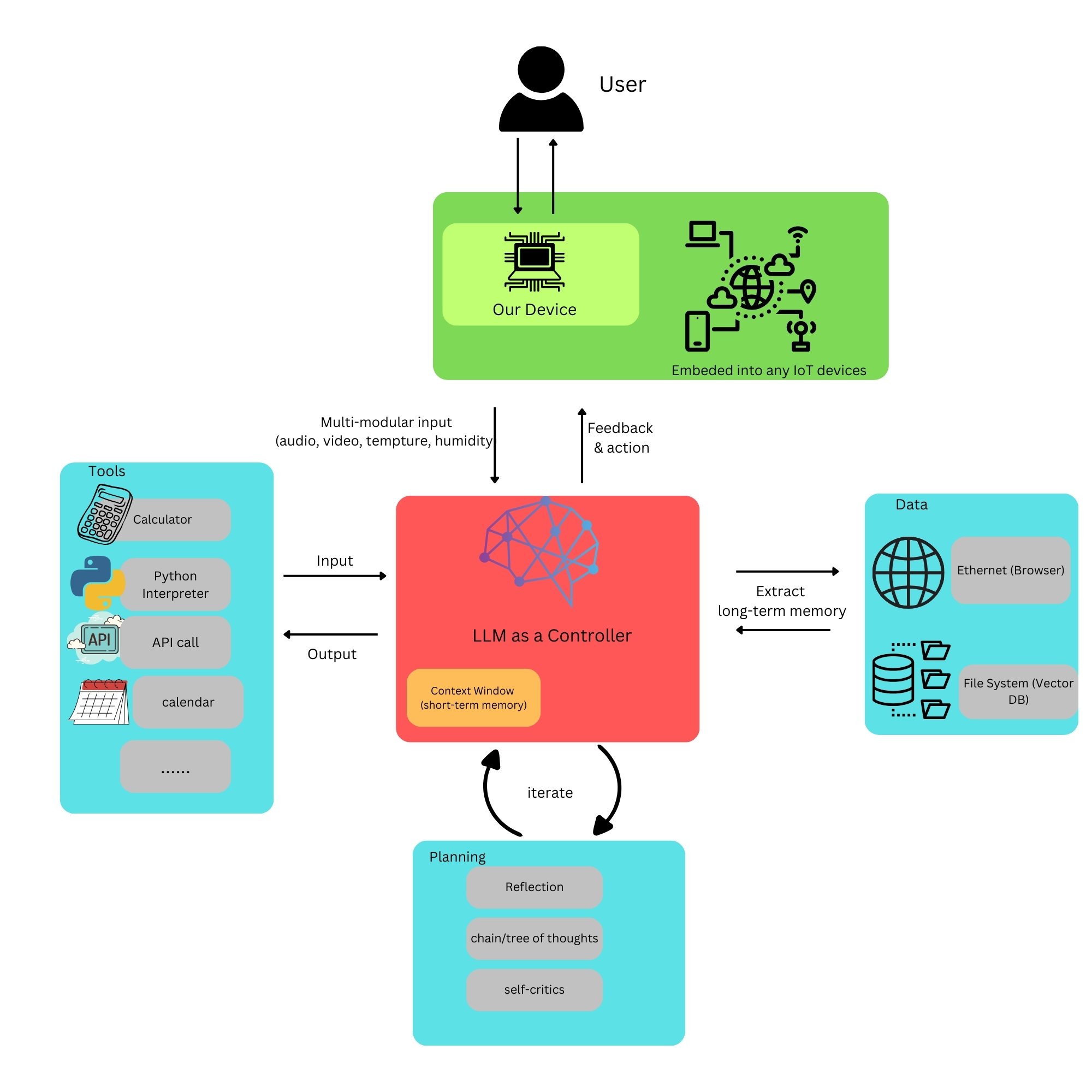}
    \caption{Framework of proposed design.\label{fig1}}  
\end{figure}

As shown in Fig. \ref{fig1}, our proposed framework consists of five major components: input edge device, LLM controller, third-party tools API, database, and task planning library.

The input edge device is close to the user end as an interactive device and can be integrated into different application scenarios such as smart homes, intelligent cities, etc. We design input signal processes for audio, video, and other multiple-dimensional data from environment sensors. The preprocess algorithms are deployed to effectively translate the raw input to a format that can be easily handled with LLM. Moreover, compressed LLM will be introduced at the edge with cache assistance to accelerate the response time and preserve sensitive practice at the user end.

The back-end LLM controller components act as the brains of the whole system and are deployed on remote servers. We will first utilize the existing LLM as ChatGPT with prompt engineering design and training our own LLM model for more specific scenario needs and precise prediction.

Our framework also includes APIs that can be integrated with third-party tools to satisfy user demands. This provides users with the flexibility to explore more powerful use cases.

We maintain our database for users' profiles and high-level features from the output of edge models for LLM improvements.
We also maintain a planning library with different use case templates; this can help LLM controllers to be quickly and easily adapt to similar common needs.

\section{Edge Device Implement}
\subsection{Major Hardware Components for Our Framework}

\noindent\textbf{Multi-modal Sensor Integration:}
Our device is uniquely designed to handle multi-modal information, making it adept at processing inputs from various dimensions. This capability is essential for understanding and interacting with the environment in a comprehensive manner.
At the core of our sensor suite are audio and video sensors. The microphone captures high-fidelity audio signals, enabling the device to process voice commands and ambient sounds. Similarly, the camera component is crucial for visual data acquisition, allowing the device to interpret visual cues and engage in image-based tasks.
To further cater to a wide range of use cases, the device is equipped with additional sensors like temperature, humidity, motion, and infrared human sensors. These sensors enrich the device's understanding of its surroundings, making it highly versatile in various environments, from smart homes to industrial settings.


\noindent \textbf{Offline Awakening Processor:}
Effective power management is a critical aspect of our hardware design. We incorporate advanced power-saving technologies and optimize each module to ensure that the device runs efficiently, maximizing battery life and reducing energy consumption.
The device incorporates an offline awakening feature to enhance battery efficiency. Using the ASR PRO technology, the device can be awakened through a wake-up word and return to a low-power sleep mode when not in active use. This functionality not only conserves energy but also ensures the device is always ready for user interaction.

\noindent \textbf{ Edge LLM model with Wireless Module:}
Despite advancements in running LLMs remotely, our device opts for a smaller, less resource-intensive LLM model for quick local response. By applying quantization and memory-reduction techniques \cite{streamllm}, we strike a balance between performance and efficiency. Moreover, to better facilitate the integration of information processing and response feedback, our edge devices integrated a WiFi module for accessing real-time information from the internet, thus keeping the device's responses and actions timely and relevant.
It also has a Bluetooth module to enable the device to communicate with local devices within its range. It allows for control and interaction with various Bluetooth-enabled devices, facilitating tasks such as home automation, data transfer, and remote device management.

\noindent\textbf{Overall Design Considerations:}
The overall hardware architecture is designed to be compact yet powerful. Each component, from sensors to processors, is carefully selected and integrated to ensure optimal performance without compromising the device's form factor.
The design also takes into account the need for scalability and upgradability. As technology evolves, our hardware can be easily updated or scaled to accommodate new sensors, improved processors, and other advancements, ensuring the device remains relevant and effective over time.

Our device's hardware design harmoniously blends multimodal sensor capabilities, efficient processing units, and robust connectivity options. This design not only addresses the current needs of users interacting with LLMs but also anticipates future advancements, ensuring longevity and relevance in a rapidly evolving technological landscape.

\subsection{Input Process }
\noindent \textbf{Audio Input and ASR Model Integration}
The primary input modality in our system is audio, which necessitates a robust and accurate translation into text. This translation is accomplished through an Automatic Speech Recognition (ASR) model. The ASR model serves as the cornerstone for interpreting user commands and queries, making its efficiency and accuracy critical for the overall performance of the device.
To ensure that the audio input is translated into a meaningful and accurate text message, we employ advanced noise-reduction algorithms. These algorithms are designed to filter out background noise and focus on the primary audio signal, which is essential for clear and precise voice recognition, especially in environments with varying noise levels.

The system utilizes state-of-the-art signal processing techniques to enhance the clarity and quality of the audio input before it reaches the ASR model. This preprocessing step is vital for mitigating issues such as echo, distortion, and varying speech volumes.
The audio input is sampled at a high frequency to capture a wide range of human speech nuances. A higher sampling rate ensures that the audio signal is accurately represented and processed, leading to more accurate speech recognition results.

\noindent \textbf{Multimodal Input Processing}
In addition to audio, our system is capable of processing multimodal inputs, which include visual (from the camera) and various environmental data (from other sensors).
The visual input from the camera is processed using image recognition algorithms. This allows the system to interpret visual cues and integrate this information with audio inputs for a more comprehensive understanding of the user's environment and intentions.
The camera’s frame rate is optimized to balance between providing smooth visual input and minimizing processing load. This ensures efficient and timely processing of visual data without overburdening the system.

Inputs from environmental sensors like temperature, humidity, and motion are processed in real-time. This data provides contextual information that can enhance the system’s responses and actions, making them more relevant to the current environmental conditions.
 The sampling rate for environmental sensors is set based on the nature of the data they collect. For instance, temperature and humidity sensors might not require as frequent sampling as motion or infrared sensors, which need to respond promptly to changes in the environment.

To summarize, the input process in our system is a carefully orchestrated procedure that prioritizes accuracy, efficiency, and context-awareness. By integrating a sophisticated ASR model with advanced noise reduction and processing multimodal inputs, the system is well-equipped to understand and respond to a diverse range of user commands and environmental stimuli. The optimization of sampling frequencies across different modalities further ensures that the system remains responsive and accurate, providing a seamless and intuitive user experience.

\subsection{Detailed Audio Input Processing}
Voice Activity Detection (VAD) is a crucial element in our audio input process. VAD is responsible for identifying the presence of human speech within the audio stream. This step is essential for distinguishing between speech and non-speech segments, allowing the system to focus processing resources on relevant audio data, thereby improving efficiency and reducing computational load.
Our VAD algorithm is designed to be highly sensitive to variations in speech, ensuring accurate detection even in noisy environments. It dynamically adjusts its parameters based on the ambient noise levels to maintain consistent performance.

Moreover, Acoustic echoes, caused by the audio output from the device being captured by the microphone, can significantly degrade the quality of the audio input. This poses a challenge for accurate speech recognition. To counteract this, our system implements advanced Acoustic Echo Cancellation (AEC) technology. AEC works by identifying and eliminating the echo from the audio signal, ensuring that the ASR receives a clean input, free from distortions caused by feedback.

Our system employs sophisticated denoising algorithms to filter out background noise. In addition, de-reverberation techniques are used to reduce the impact of sound reflections, especially in enclosed spaces. These processes are crucial for enhancing speech clarity and intelligibility.
Both the denoising and de-reverberation algorithms are adaptive, meaning they adjust their parameters in real-time based on the characteristics of the input signal, ensuring optimal performance in a variety of acoustic environments.

The heart of our audio input process is the Automatic Speech Recognition model. This model converts spoken language into text, enabling the device to understand and act upon user commands.
The ASR model works in tandem with the previously mentioned audio preprocessing steps. By receiving a cleaner, more focused audio input, the ASR can perform more accurately, leading to a better user experience.
The overall workflow of our audio input processing is designed to be seamless and efficient. It begins with the activation of the VAD, followed by AEC to eliminate any echo, then denoising and de-reverberation to clarify the signal, and finally, the refined audio is fed into the ASR model. This process is continuously optimized based on ongoing feedback and learning from user interactions. This ensures that the system not only maintains its performance over time but also adapts to new challenges and environments.

\begin{figure}
    \centering
    \includegraphics[width=\linewidth]{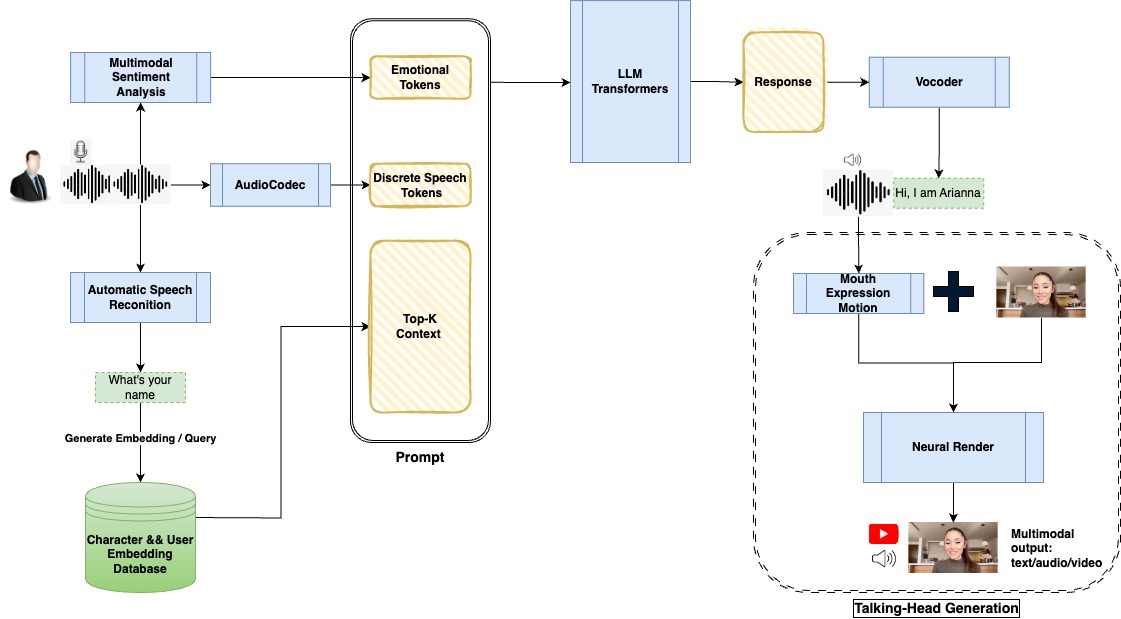}
    \caption{Information process flow from edge to core.}
    \label{fig2}
\end{figure}

The detailed audio input processing in our system is an intricate combination of various technologies, each contributing to the goal of providing clear, accurate, and efficient voice recognition, see Fig. \ref{fig2}. From voice activity detection to automatic speech recognition, every step is meticulously designed and integrated, ensuring that the system can understand and respond to user commands effectively, even in challenging acoustic environments. This attention to detail in audio processing is a cornerstone of our device’s ability to offer a seamless and intuitive user experience.




\subsection{Local Caching}
The inclusion of a local cache in our edge device design represents a significant enhancement in the efficiency and responsiveness of Large Language Model (LLM) interactions. This local cache system is strategically implemented to store the most frequently asked questions, their corresponding answers, and actions. The primary objective of this feature is to facilitate rapid response times and reduce reliance on continuous cloud connectivity, which can be a limiting factor in terms of latency and bandwidth usage.

\noindent \textbf{Functionality and Benefits:}
 By caching the most common queries and actions, the device can offer instant responses to frequent requests. This is particularly beneficial in scenarios where speed is crucial, such as in emergency situations or in fast-paced work environments.
 Local caching significantly decreases the need for constant data transmission between the edge device and the cloud. This reduction in data exchange not only conserves bandwidth but also minimizes potential latency issues, especially in areas with limited or unstable internet connectivity.
Storing data locally minimizes the amount of sensitive information transmitted to and from the cloud, thus enhancing user privacy. This feature is particularly important for users who are cautious about their data privacy and prefer to keep their interactions confined to the device.

The local cache is not static; it evolves based on user interactions. This means that the cache is continuously updated with customized data, reflecting the unique needs and usage patterns of each user. Such customization ensures that the cache remains relevant and effective over time.
 The cache updating mechanism is designed to be dynamic, ensuring that the stored data remains current and useful. The device will periodically review the cache contents, adding new frequently asked questions and actions while removing outdated or less frequently used data.
The device’s local cache system is integrated with an adaptive learning algorithm. This allows the cache to not only store data but also learn from user interactions, thereby improving its predictive capabilities and ensuring that the most relevant information is readily available.

 The design of the local cache focuses on efficient storage management. This involves optimizing the data structure to ensure that the cache can store a significant amount of data without overwhelming the device’s memory and processing capabilities.

Security is a paramount concern in the design of the local cache. Robust encryption and secure data handling protocols are implemented to protect the stored data from unauthorized access or breaches.

\noindent \textbf{Challenges and Considerations:}
 Balancing the cache size with the device’s hardware limitations is a key challenge. It is crucial to ensure that the cache is large enough to be effective but not so large that it hampers the device’s overall performance.

 Ensuring the relevance and accuracy of cached data over time is another important consideration. Mechanisms must be in place to regularly validate and update the cache to maintain its effectiveness and reliability.

The local cache feature in our edge device design plays a critical role in enhancing the responsiveness and efficiency of LLM interactions. By intelligently managing frequently used data, the device not only offers quicker responses but also provides a more personalized and privacy-conscious user experience. The ongoing challenge lies in effectively managing this cache to ensure it continues to meet the evolving needs of the users while maintaining the device’s performance and security standards.

\subsection{LLM as a Controller}

In our system, the Large Language Model (LLM) functions akin to a brain, orchestrating various components and processes to accomplish complex tasks. This central role of the LLM is pivotal, as it not only processes input data but also makes informed decisions, directs actions, and manages interactions between different modules of the device.

 The LLM receives and interprets data from the array of sensors and input mechanisms in the device. By analyzing audio, visual, and environmental sensor data, it gains a comprehensive understanding of the user's context and intent. This holistic approach allows the LLM to respond more accurately and appropriately to user requests.

A crucial feature of the LLM in our system is its ability to access and search the internet. This capability allows it to fetch real-time information, reference up-to-date data, and incorporate external knowledge into its responses. This is particularly beneficial for tasks that require current information or specialized knowledge beyond the device's local storage.

The LLM serves as the command center for executing complex tasks. Whether it’s turning on smart home devices, setting reminders, or providing detailed explanations on various topics, the LLM evaluates the best course of action and executes it efficiently.

 In performing tasks, the LLM seamlessly switches between utilizing local resources (like stored data and connected devices) and remote resources (like cloud services and online databases). This flexibility ensures that the system uses the most appropriate and efficient means for each task.

 Once a task is completed or a query is resolved, the LLM formulates a response. These responses are not just text-based but can also be in the form of actions, like adjusting the thermostat or displaying visual information on a connected screen.
The LLM is designed to be context-aware, tailoring its responses and actions to the specific situation and user preferences. This adaptability enhances the user experience, making interactions more intuitive and personalized.

An essential aspect of the LLM in our system is its ability to learn and adapt over time. By analyzing user interactions and feedback, the LLM continuously improves its performance, becoming more efficient and effective in handling various tasks and scenarios.

As the LLM handles sensitive data and controls various aspects of the user's environment, robust privacy and security measures are integral. The system ensures that all interactions and data handling adhere to strict security protocols, safeguarding user privacy.

The LLM in our system acts as a sophisticated controller, leveraging its advanced capabilities to manage a wide range of tasks and interactions. Its ability to integrate multimodal inputs, access vast online resources, and adapt to user needs positions it as an invaluable asset in our system. This approach not only enhances the functionality and efficiency of the device but also elevates the overall user experience, making it more responsive, intuitive, and secure.

\subsection{Output Feedback}

\noindent\textbf{Integration of Cloud-Based LLM and Local Device:}
In our system, the output feedback mechanism plays a critical role in bridging the capabilities of the cloud-based Large Language Model (LLM) with the functional execution of the local device. This seamless integration ensures that the intelligence and processing power of the LLM are effectively translated into actionable responses and tasks performed by the local device.

The LLM, residing in the cloud, processes and interprets the user's requests and the data gathered from various inputs. Once a decision or response is formulated, this information is communicated back to the local device. This communication is designed to be swift and secure, ensuring that the feedback is delivered without significant latency and with data integrity maintained.

 Upon receiving feedback from the LLM, the local device acts as an executor, translating the instructions into physical actions. This includes controlling connected devices such as smart home appliances, adjusting settings, or displaying information on connected screens. The device's ability to interact with a wide range of connected technologies makes it a versatile tool for executing a variety of tasks.

 The feedback from the LLM is not limited to text-based responses. Depending on the nature of the task and the user’s preferences, the output can be in various forms – from spoken words (through connected speakers) to visual displays or even physical actions enacted by connected devices.

\noindent \textbf{Feedback Adaptation and Learning:}
The LLM tailors its responses based on the context it gathers from the local device’s inputs. This context-aware approach ensures that the feedback is not only accurate but also relevant to the user's current environment and situation.

 The system is designed to learn from each interaction. By analyzing the outcomes of its feedback and the user's reactions, the LLM fine-tunes its future responses. This adaptive learning process continually enhances the relevance and effectiveness of the feedback provided.

\noindent\textbf{Local Device’s Role in Feedback Loop:}
 The local device is responsible for the real-time implementation of the LLM’s feedback. Whether it’s a simple command or a complex series of actions, the local device ensures that the feedback is executed accurately and efficiently.

The system also allows for user feedback on the actions taken. This feedback is crucial for the system’s learning process, as it provides direct input on user satisfaction and preferences, which is used to refine future responses and actions.

The output feedback mechanism is a vital component of our system, enabling the cloud-based LLM to effectively communicate and control actions through the local device. This system not only ensures efficient and accurate task execution but also adapts and evolves to better suit the user’s needs, all while maintaining high standards of security and privacy.

\section{Future Work and Challenges}
\subsection{Future Work}
\noindent\textbf{Advancing Hardware Integration for LLMs:}
  Future work should develop scalable and efficient hardware designs that seamlessly integrate with LLMs. This involves creating customized circuit boards and processing units tailored to the computational demands of LLMs. Emphasis should be placed on energy efficiency and heat management to enable widespread adoption in various environments, from personal devices to industrial settings.
 Enhancing edge computing capabilities is crucial for deploying LLMs in remote and resource-constrained environments. Future research should aim at miniaturizing hardware while maximizing computational power, ensuring that LLMs can operate effectively in edge devices without compromising speed or accuracy. For example, we may exploit the quantization technique \cite{bit} and apply the models with fewer parameters \cite{mobilellm}.

\noindent\textbf{Improving Multimodal Data Processing:}
 Advanced Multimodal Sensor Integration: Integrating advanced multimodal sensors in LLM hardware will be a crucial area of future research. This includes developing sensors that can accurately capture and process diverse data types (e.g., visual, auditory, tactile) in real-time, enabling LLMs to provide more context-aware and responsive interactions.
 Developing algorithms capable of real-time multimodal data processing is essential. This requires advancements in parallel processing techniques and the optimization of algorithms for speed and accuracy, ensuring that LLMs can respond to complex scenarios promptly and effectively.

\noindent\textbf{Enhancing User Interaction and Accessibility:}
 Future research should focus on enhancing the personalization capabilities of LLM-integrated devices. This includes adaptive learning algorithms that can tailor interactions based on individual user preferences, histories, and behaviors, thus providing a more intuitive and user-centric experience.
 Accessibility Improvements: Improving accessibility for a broader range of users, including those with disabilities, is a critical area of future work. This involves designing hardware and interfaces adaptable to various accessibility needs, ensuring that LLM technologies are inclusive and beneficial to all users.

\noindent\textbf{Processing Stochastic Data and Enhancing Performance:}
  Future work must address the challenges posed by real-world data's inherently noisy and stochastic nature \cite{Anomaly}, which involves developing advanced noise reduction techniques \cite{ensemble_p}, stochastic data modeling, and adaptive filtering methods to effectively differentiate between meaningful signals and noise across various data types. Simultaneously, research should focus on leveraging ensemble methods \cite{ensemble} to achieve better performance in handling complex, real-world scenarios . This approach includes integrating multiple LLM or specialized models, developing advanced ensemble learning algorithms tailored for LLM applications, implementing dynamic model selection based on specific contexts, and incorporating uncertainty quantification methods. By combining robust stochastic data processing with powerful ensemble techniques, future LLM-integrated hardware can significantly improve its performance, reliability, and adaptability in diverse and noisy environments, making these systems more effective in real-world applications.

\subsection{Overcoming Implementation Challenges}
 Addressing scalability and maintenance challenges is critical for the widespread adoption of LLM-integrated hardware. This includes developing modular designs that can be easily upgraded and maintained and scalable deployment strategies that accommodate growing data and processing demands.
 Ensuring interoperability with existing technologies and platforms is essential for the seamless integration of LLMs into current systems. This involves developing hardware and software interfaces compatible with various devices and applications, facilitating easier adoption and integration.

\section{Conclusion}
In this paper, we have explored the burgeoning field of integrating LLMs with advanced hardware, particularly focusing on the development of a General-Purpose Device for Interaction with LLMs. Our journey began with an examination of the current landscape, where virtual assistants and LLMs have significantly altered how we interact with technology. We delved into the existing body of work, highlighting key advancements and innovations that have set the stage for this exciting new era of intelligent hardware.

Our exploration revealed that while significant progress has been made in the realm of LLMs, there remains a substantial gap in hardware capabilities, particularly in terms of scalability, efficiency, affordability, and multimodal integration. This gap presents both a challenge and an opportunity for researchers and developers in the field. The potential of LLMs can only be fully realized through hardware that is not just powerful but also versatile, energy-efficient, and capable of handling the complex demands of modern computation.

The proposed general-purpose device is a step towards bridging this gap. By focusing on scalability, multimodal data processing, user interaction, and privacy concerns, this device aims to provide a robust platform for LLM integration across various applications. Our discussion on future work and challenges underscores the need for continuous innovation in this domain, particularly in areas such as edge computing, real-time data processing, penalization, and ethical AI.

As we conclude, it is evident that the journey towards fully realizing the potential of LLMs in practical applications is still in its infancy. The integration of LLMs with dedicated hardware presents a unique set of challenges and opportunities that require a collaborative effort from researchers, developers, industry experts, and ethical committees. The future of this field is not only about technological advancement but also about responsible and inclusive innovation.

%
%
%
\bibliographystyle{IEEEtran} 
\bibliography{main}






\end{document}